\documentclass[aps,prl,twocolumn,showpacs,superscriptaddress,10pt]{revtex4}

\usepackage{latexsym}
\usepackage{graphicx}
\usepackage{amsmath}
\usepackage{amsfonts}
\usepackage{amssymb}
\usepackage{amsthm}
\usepackage{amsmath}
\usepackage{times}
\usepackage[colorlinks,citecolor=blue,linkcolor=blue]{hyperref}
\usepackage{amsbsy}
\usepackage{float}
\usepackage{epstopdf}
\usepackage{hyperref}
\usepackage{color}
\DeclareGraphicsExtensions{.pdf,.eps,.png,.jpg,.mps}

\begin{document}
\title{Predicting market instability: New dynamics between volume and volatility}
\author{Zeyu Zheng}
\affiliation{Department of Physics and Centre for Computational Science
  and Engineering, National University of Singapore,
Singapore 117542, Republic of Singapore}

\author{Zhi Qiao}
\affiliation{Department of Physics and Centre for Computational Science
  and Engineering, National University of Singapore,
Singapore 117542, Republic of Singapore}
\affiliation{NUS Graduate School for Integrative Sciences and
  Engineering, National University of Singapore,
Singapore 117456, Republic of Singapore}

\author{Joel N. Tenenbaum}
\affiliation{Center for Polymer Studies and Department of Physics, Boston
University, Boston, MA 02215, USA}

\author{H.~Eugene~Stanley}
\affiliation{Center for Polymer Studies and Department of Physics, Boston
University, Boston, MA 02215, USA}

\author{Baowen Li}
\affiliation{Department of Physics and Centre for Computational Science
  and Engineering, National University of Singapore,
Singapore 117542, Republic of Singapore}
\affiliation{NUS Graduate School for Integrative Sciences and
  Engineering, National University of Singapore,
Singapore 117456, Republic of Singapore}

\date{\today --- zqtl12Aug2.tex}

\begin{abstract}

Econophysics and econometrics agree that there is a correlation between
volume and volatility in a time series. Using empirical data and their
distributions, we further investigate this correlation and discover new
ways that volatility and volume interact, particularly when the levels
of both are high.  We find that the distribution of the
volume-conditional volatility is well fit by a power-law function with
an exponential cutoff.  We find that the volume-conditional volatility
distribution scales with volume, and collapses these distributions to a
single curve.  We exploit the characteristics of the volume-volatility
scatterplot to find a strong correlation between logarithmic volume and
a quantity we define as local maximum volatility (LMV), which indicates
the largest volatility observed in a given range of trading volumes.
This finding supports our empirical analysis showing that volume is an
excellent predictor of the maximum value of volatility for both same-day
and near-future time periods.  We also use a joint conditional
probability that includes both volatility and volume to demonstrate that
invoking both allows us to better predict the largest next-day
volatility than invoking either one alone.


\end{abstract}

\maketitle

\section{Introduction}

It is common knowledge among investors that trading volume is strongly
connected to price change \cite{KK,KKK}, and the price-volume
relationship in financial markets has been a popular research topic for
economists for a long time. Although studies agree that there is a
correlation between absolute price change (volatility) and trading
volume \cite{Ane}, many indicate that the correlation is weak
\cite{Karpoff} and their analyses of time-lag correlations produce a
variety of contradictory results
\cite{Lamoureux,Plerou2,Ying,Crouch1,Rogalski,Whaley,Tauchen1,Gabaix1}.
The subtleties of the relationship between volume and volatility remain
unclear \cite{Lobato} and disagreement persists. For example, Brailsford
et al.~report a significant cross-correlation between overnight return
and trading volume \cite{Brailsford}. Brooks indicates that including
lagged volume may lead to modest improvements in forecasting performance
\cite{Brooks} while Clark shows a nearly parabolic functional
relationship between volume and volatility \cite{Clark}, and a popular
model developed by Clark holds that volatility could be modeled as a
subordinated random process, in which volume, insofar as it affects
trading times, accounts for the majority of observed volatility
clustering and leptokurtosis (i.e., heavy tails).  On the other hand,
several studies report that volume is only nominally useful in
predicting volatility. Koulakiotis et al.~report a negative relationship
between volatility and trading volume \cite{Koulakiotis}. Lamoureux and
Lastrapes show that ARCH effects tend to disappear (i.e., volatility
persistence is lost) when volume is included in the variance equation
\cite{Lamoureux}.  Sharma et al.~even suggest that price returns of the
NYSE are best described by the GARCH model in the absence of volume as a
mixing variable \cite{Sharma}. Recently, Gillemot et al.~demonstrated
that the subordinated random process developed by Clark accounts for, at
most, only a small fraction of observed volatility clustering and
leptokurtosis \cite{Gillemot}.



In order to uncover the underlying relationship between volume and
volatility, we focus on the most fundamental features of these two
quantities, starting by examining the probability density function (pdf)
of each, as well as the volume-conditional pdf of volatility in our
dataset.  Based on these elementary analyses, we show that the pdf for
volume-conditional volatility is actually invariant under volume change
when the units of volatility are scaled appropriately, in close
connection to similar work carried by out Yamasaki et al., who reported
a universal scaling function for return intervals of
volatility \cite{Yamasaki}.  We then propose a new probability density
function which links the occurrence of volatility and volume.  We
further investigate the highest portion of volatility distribution in
certain volume regimes and propose a quantity we refer to as ``local
maximum volatility'' (LMV), which we show is closely related both to a
given day's volume, as well as the volume of days previous.

\section{Data and Methods}

We analyze the 30 stocks comprising the Dow Jones Industrial Average,
using daily values from the 17-year time period from April 1990 to June
2007, for a total of 130,410 data points. We avoid data after June 2007
due to the potential for high non-stationarity in the volume time series
associated with the world financial crisis, although further analysis
indicates that our results do not change when post-2007 data is
included.

For each of the 30 stocks $i$, we calculate the daily logarithmic
change, commonly referred to as the {\it return}, of price $p_i(t)$
\begin{equation}
R_i(t)  \equiv ln p_i(t) - ln p_i(t-1),
\label{Return}
\end{equation}
and also the daily normalized logarithmic trading volume $\tilde
Q_i(t)$, calculated from the trading volume $Q_i(t)$ as 
\begin{equation}
\tilde  Q_i(t) \equiv ln Q_i(t)-Y_i,
\label{Volume}
\end{equation}
for a given stock $i$, where $Y_i$ represents a least-squares linear fit
of $\ln Q_i$ \cite{Chambers}, which removes the global trend over the 
entire 17-year period. For each different stock, we define the
normalized volatility $g_i(t)$ and normalized logarithmic volume
$v_i(t)$ from the raw returns and raw logarithmic volume by 
\begin{equation}
g_i(t)  \equiv  {\big|}\frac{R_i(t) - \langle R_i(t)\rangle}{\sigma_R }{\big| }
\label{nr}
\end{equation}
and
\begin{equation}
v_i(t)  \equiv \frac{\tilde Q_i(t) -\langle \tilde Q_i(t) \rangle}{\sigma_{\tilde Q}},
\label{nv}
\end{equation}
where $\langle\cdots\rangle$ denotes a time average over the period
studied. Here $\sigma_R =\sqrt{ \langle R^2\rangle-{\langle
    R\rangle}^2}$ and $\sigma_{\tilde Q} =\sqrt{
  \langle\tilde{Q^2}\rangle-{\langle \tilde{Q}\rangle}^2 }$ are the
standard deviations of $R(t)$ and $\tilde Q(t)$, respectively. Note that
the volatility is expressed in terms of absolute value while the
logarithmic volume can be both positive and negative. In this paper, the
volume indicates the normalized logarithmic volume $v_i(t)$, and
volatility indicates the normalized volatility $g_i(t)$.

\section{Analysis}

\begin{figure}
\centering \includegraphics[width=0.5\textwidth]{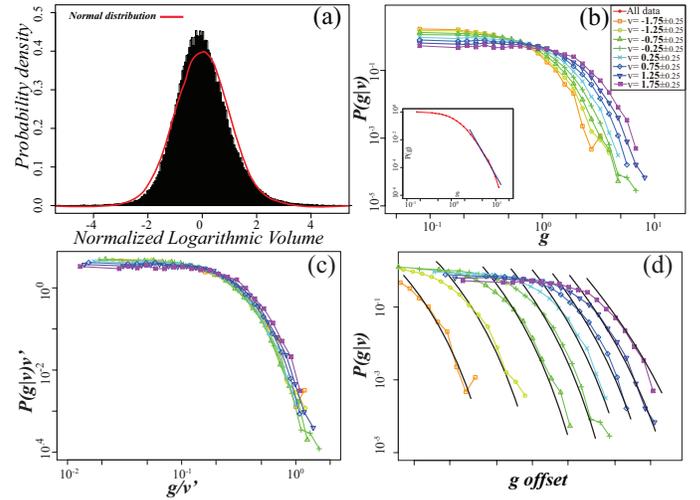}\\
\caption{Range of volumes in the dataset, as well as probability density
  function (pdf) $P(g$) and conditional pdf $P(g|v)$, demonstrating that
  all conditional volatility distributions behave the same in scaled
  units. (a) Pdf of normalized logarithmic volume $v$ (Eq.~\ref{nv}) for
  the 30 Dow Jones Industrial Average stocks for the period from April
  1990 to June 2007, well fit by a normal curve (solid line). (b)
  Conditional pdf showing the distribution of volatility given a
  specified volume range. Inset: unconditioned pdf of volatility,
  roughly fit by a power-law distribution. (c) When the curves in (b)
  are plotted in scaled parameter of $v'$, where $v' \equiv v+4.5$, data
  for all volumes collapse onto the same curve. (d) Same as (b), but
  shifted horizontally for better visibility with tails given power-law
  fits with exponential cut offs described by Eqs.~\ref{POW} and
  \ref{cut}.}
\label{1}
\end{figure}

We begin by examining the probability density function (pdf) of the
normalized logarithmic trading volume, which we find in Fig.~\ref{1}(a)
to be in excellent agreement with a unit Gaussian. The normal curve is
often a null model for various econometric quantities. For example, Wang
et al.~\cite{Wang} have shown that a normal curve is also a good fit for
trading values. However, the pdf of volatility is widely known to be
more leptokurtic (i.e., fat-tailed) than a normal fit, which we show in
the inset picture in Fig.~\ref{1}(b) as a log-log plot. The solid red
line is the pdf of volatility, the tail of which we observe roughly
matches a power-law distribution, as was pointed out in
Ref.~\cite{Wang2}. We also find the distribution of returns to be
leptokurtic as well, being better fit by a Laplace distribution than a
Gaussian, in agreement with work by Podobonik et al.~on NYSE
stocks \cite{Podobnik2009}.

The tendency of trading volume and price change to move together has
important implications in the prediction of financial risk. Recent
studies have revealed the long-term cross-correlation of volume changes
with price changes \cite{Podobnik}, and also the positive correlation of
price changes with volume \cite{Gallant,Karpoff}. As the absolute value
of return, volatility should be a better indicator for market
fluctuation and so we investigate the pdf of volatility given a
specified volume.  As shown in Fig.~\ref{1}(b), the conditional
volatility distributions for various volumes seem very similar, which
leads us to search for scaling features that unify these distributions.
We draw inspiration from the work of Yamasaki et al., who analyzed the
distribution of return intervals $\tau$ between volatilities larger than
a specified threshold $q$ \cite{Yamasaki}.  They found that the
distributions for different $q$ across seven stocks and currencies all
collapsed to a single curve when plotted in units scaled by the mean
return interval, dependent on $q$.  We investigate here whether a
similar scaling parameter exists that could unify these distributions.
This scaling parameter should incorporate volume dependence the same way
$\tau$ incorporates $q$ dependence in Yamasaki's work.

Redrawing the conditional volatility distributions using the scale
parameter $v'$, where $v' \equiv v + 4.5$, results in all conditional
distributions collapsing onto the same curve, regardless of the value of
volume, as shown in Fig.~\ref{1}(c), meaning that all conditional
volatility distributions are unified, differing only by a factor of the
volume chosen, very similar to Yamasaki's findings on volatility return
intervals.  We have chosen the offset in a volume of $4.5$ to avoid
singularities and unphysical values, since normalized volume as defined
in Eq.~\ref{nv} can be a non-positive quantity.

We next investigate what unified pdf these distributions follow. In
Fig.~\ref{1}(d) the volume-conditional pdfs are offset for better
visibility. We notice that the tails of these distributions are too
curved to fit power-laws.  After investigating such distributions as
log-normal and stretched exponential, we find the best fit using
power-law distributions with exponential cutoffs.  Thus the distribution
of volatility given a certain value of volume should be


\begin{equation}\label{POW}
P(g|v)~\sim~ g^{-\xi}e^{-\varsigma g}.
\end{equation}

However, as Fig.~\ref{1}(c) shows, the above pdf can be scaled in
$v'$($v' \equiv v + 4.5$), which leads us to add volume as a variable of
the conditional volatility distribution function. Thus we assume $\xi =
\alpha v+a$ and $\varsigma = \beta v+b$, making Eq.~\ref{POW} 


\begin{equation}\label{cut}
P(g,v|v)~\sim~ g^{-(\alpha v+a)}e^{-(\beta v+b)g}.
\end{equation}

\begin{figure}
\centering \includegraphics[width=0.6\textwidth]{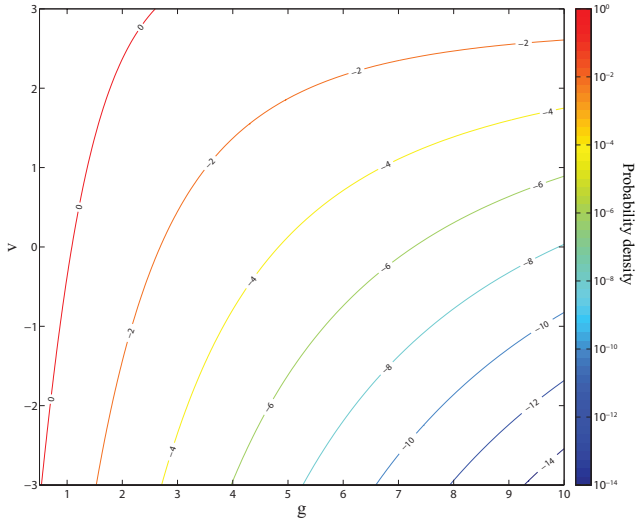}\\
\caption{Fit of the joint distribution of volatility and volume.  Using
  fitted values for $\alpha$, $\beta$, $a$, $b$, we show the contour
  plot for the probability density function of Eq.~\ref{cut} to fit the
  data seen in Fig.~\ref{1}(b). We show that either decreasing
  $g$(volatility) or increasing $v$(volume) results in monotonic
  increases in the probability density.  Higher volatilities are far
  more localized in their range of volumes than low volatilities,
  leading to the possibility that higher volatilities may be predicted
  from volume.}
\label{2}
\end{figure}

Using a maximum likelihood estimation for the data shown in
Fig.~\ref{1}(b), we find $\alpha=0.4$, $\beta=-1.23$, $a-2.5$, and
$b=3$.  We draw a contour plot using these parameters with Eq.~\ref{cut}
in Fig.~\ref{2}, showing that $g$ (volatility) and $v$ (volume) increase
concurrently given a certain probability density value. Specifically, we
note that while low volatilities can occur over the entire range of
volumes fairly regularly, higher volatilities have a strong tendency to
occur only with larger volumes, meaning that high volatilities may be
predictable from volume, although low volatilities cannot.


As a consequence, we restrict our analysis to the days comprising the
largest portion of volatilities---which is appropriate, given that days
of high volatility are the ones of greatest interest to traders and
market researchers. To do this we introduce the quantity ``local maximum
volatility'' (LMV), which, because it is closely related both to a given
day's volume and the volume of previous days, allows the possibility of
making predictive statements.

We define the LMV parameter, denoted by $g_{\rm LM}$, by partitioning
the range of observed trading volumes into bins $u_1, u_2,u_3,\ldots
u_n$. Then
\begin{equation}\label{LMV}
g_{\rm LM}^j\equiv \max(\{g_t\}) \forall t \mid v_t \in u_j.
\end{equation}

\begin{figure}
\centering \includegraphics[width=0.5\textwidth]{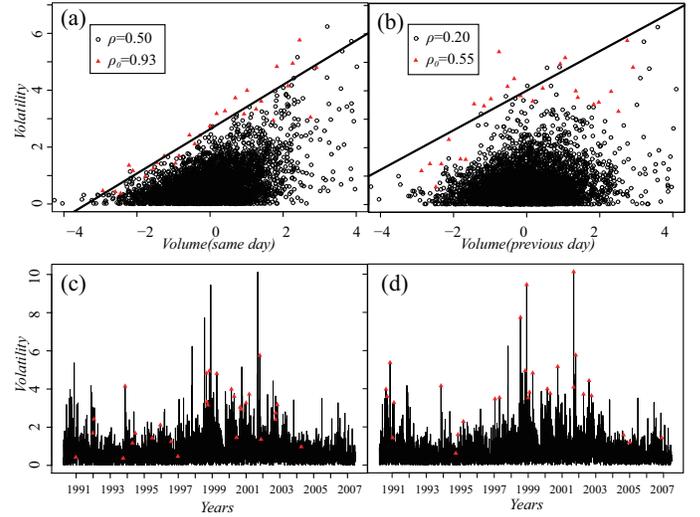}\\
\caption{While volatility is not highly correlated to volume, LMV is
  highly correlated to both today's volume and to yesterday's volume
  (linear fits for LMV shown). LMV days occur throughout the period
  which we study. Shown is a scatter plot of volatility vs. volume for
  the example of The Boeing Company (BA): (a) Volatility $g(t)$ vs.
  normalized logarithmic volume $v(t)$, (b) volatility $g(t)$
  vs. normalized logarithmic volume the day before, $v(t-1)$. The red
  solid triangles depict the largest values in each bin of g (from -3 to
  3 we delineate 30 bins evenly). $\rho_0$ is the correlation
  coefficient between logarithmic volume and LMV(Eq.~\ref{LMV}), while
  $\rho$ represents correlation coefficient between logarithmic volume
  and volatility.  The volatility time series and LMV (red triangle) are
  shown for LMV based on (c) concurrent volume and (d) previous day's
  volume.  }

\label{3} 
\end{figure} 

LMV is the maximum volatility observed in a given range of trading
volumes, i.e., the volatility of the most volatile day a given trading
volume has co-occured with. Although correlation between volatility and
logarithmic volume is weak, we find that, in general, LMV and
logarithmic volume are highly correlated.  We demonstrate this in
Fig.~\ref{3}(a) using the example of the Boeing Company (BA). For BA, we
observe that while the correlation coefficient between same-day volume
and volatility is only 0.5, the correlation coefficient between volume
and LMV is 0.93.  We further investigate the correlation between
volatility and volume using the scatter plot of volatility against
volume in Fig.~\ref{3}.  A characteristic triangular shape can be seen
in both the scatter plot of (a)volatility vs. the same-day volume and
(b)volatility vs. the previous day's volume. The volume ranges used to
define LMV are delimited by defined bins as is shown in Fig.~\ref{1}(a)
(30 bins evenly divided from $-3$ to 3). As defined in Eq.~\ref{LMV}, we
use the highest volatility in each given bin. In both cases, the maximum
volatility matching a given volume is shown in red triangles and a
linear regression fit is shown in solid black, visually confirming the
calculated correlation.  Because it is possible that the volatilities
used in LMV could originate in a narrow, unusually volatile time window
(e.g., one week), and thus be giving spurious results, we investigate
the timing of the high volatility days used. Figures~\ref{3}(c) and
\ref{3}(d) show that these high volatility days do indeed occur
throughout the span of the time period under consideration, which
ensures the universal representativeness of LMV.



\begin{figure}
\centering \includegraphics[width=0.45\textwidth]{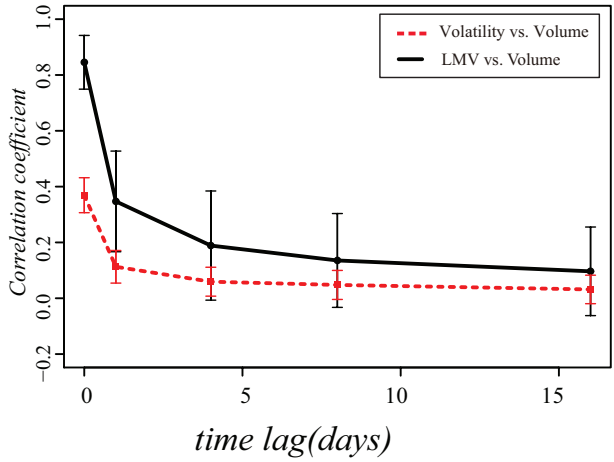}\\
\caption{There is a weak correlation between volatility and volume (red squares), though this effect quickly drops off with time lag.  LMV has a stronger correlation
with volume (black circles) throughout the range of time lags.  Shown is the  mean correlation coefficient vs. time lag for the 30 DJIA stocks.
The error bars depict $\pm$ standard deviation.  Note that the mean correlation coefficients for time-lag = 0 days, 1 day are very similar to those
found in Fig.~\ref{4}, which depicts results for only The Boeing Company (BA).}\label{4}
\end{figure}

We now generalize the analysis shown in Fig.~\ref{3} for same-day and
one-day offsets to variable time offsets up to 16 days.  Our results are
shown in Fig.~\ref{4}.  In the figure, we show the mean correlation
coefficient against time-lag for the 30 DJIA stocks.  The figure shows
that while the correlation between volume and volatility quickly drops
to zero for almost any nonzero time-lag, the correlation between volume
and LMV retains significant value ($\approx0.4$) at a one-day lag and
remains noticeable ($\approx0.2$) even with a 4-day time-lag, indicating
significant potential for predicting days of potential largest
volatility, and therefore largest risk, which is extremely important in
protecting investments during a financial crisis \cite{zheng}.


\begin{figure}[b]
\centering \includegraphics[width=0.52\textwidth]{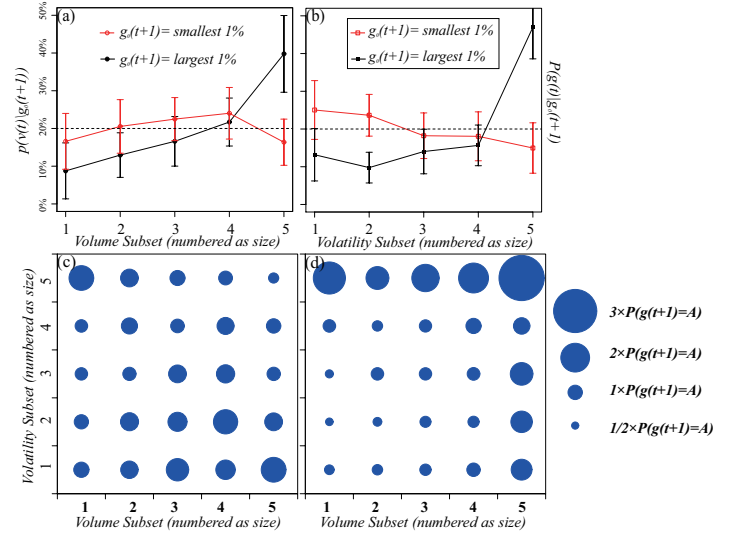}\\
\caption{Predicting power of volume and volatility. Days preceding high volatility are characterized by higher volume and higher volatility than usual.  Figures (a) and (b) show the conditional distribution of $P(v(t)|g_0(t+1))$ and $P(g(t)|g_0(t+1))$, averaged over all 30 stocks. Here $g_0(t+1)$ represents the subset containing largest (closed symbols) or smallest (open symbols) 1\% of volatility observed. The volatility $g(t)$ and normalized logarithmic volume $v(t)$ are divided into quantiles. The error bars show $\pm$ standard deviation. Figures (c) and (d) show $P(g(t+1)\in g_0|v_n(t),g_n(t))$,
the distribution by quintile (n=1...5) for v and g values of today given that tomorrow will have the (c) smallest or (d) largest 1\% of volatility, where $n$ is increasing with size and $g_0$ denotes the set of the (c) smallest 1\% of day volatilities or (d) the largest 1\% of day volatilities. The radius of the circle is proportional to $P(g(t+1)\in g_0|v_n(t),g_n(t))=1\%$, the unconditioned probability that the next day's volatility will be in the smallest or largest 1\% of volatility.}
\label{5}
\end{figure}

The fact that the possible volatility is closely tied to the same-day trading volume is intuitive, as the extent to which the price can change is a function of quantity of trading that has transpired in a given day. The connection between volume and future volatility is more interesting.   Because volatility is already widely known to correlate with its own values in the immediate future, this result may seem trivial. We later present evidence (see Fig.~\ref{5}) that our findings go beyond this obvious result, that the inclusion of volume really does add non-redundant information into the prediction scheme.  Additionally, as has been shown by Gillemot et al.~the tendency for volatility to cluster is not a simple volume effect resulting from reductions in average trading time\cite{Gillemot}.



Figure.~\ref{5}(a) and (b) show the conditional distribution of $P(v(t)|g_0(t+1))$ and $P(g(t)|g_0(t+1))$. Here $g_0(t+1)$ represents the subset that contains the highest 1\% or lowest 1\% of volatilities. Fig.~\ref{5}(a)shows the quintile distribution of the volume today, given a specified volatility tomorrow while Fig.~\ref{5}(b) shows the volatility today, given a specified volatility tomorrow. In the absence of memory effects, Figs.~\ref{5}(a) and \ref{5}(b) would be completely flat distributions, in both highest and lowest volatility cases. Instead we clearly see memory effect in highest cases. 
 20\% of volumes account for roughly 40\% of the days preceding the highest (top 1\%) volatilities. This effect is monotone across the quintiles and the most extreme example of underrepresentation being that the lowest 20\% of volumes account for approximately 10\% of the highest volatilities. For the lowest 1\% of volatilities we find no such effect. The distribution of days preceding low volatility is statistically the same as a flat distribution across volume. That
low volatility days do not have a statistically different distribution in volumes agrees with earlier observations seen in Fig.~\ref{2}
that low volatilities have a broad range across volumes, and hence are not predictable from volumes. We observe similar results when considering the distribution a day's volatility, knowing that the next day will have a particular high or particularly low volatility.  Again, days prior to high volatility are overrepresented
in the highest quintile of volatility, but days prior to low volatility have a distribution that is essentially flat across quintiles in volatility.

In summary, Fig.~\ref{5}(a) and (b) show not only that high volatility tends to follow high volatility, but also high volatility tends to follow high volume.
No such significant effects can be observed for low volatility.

Extending this analysis, we include both volume and volatility in order to better predict next-day volatility. Figure~\ref{5}(c) and (d) give
$P(g(t+1)=A|v_n(t),g_n(t))$, the distribution of the days preceding the highest or lowest 1\% of volatilities according to preceding
volatilities and volumes broken up into quintiles ($n=1\ldots 5$).  The probabilities are given in units of $P(g(t+1)=A)$, the unconditioned
probability of a defined volatility (top or bottom 1\%) day, which is equal to 1\%. Figure~\ref{5}(c) and (d) therefore divide the 1,304 data points
(1\% of all data points) into $5 \times 5 = 25$ equal-sized sets of approximately 52 points each. Figure~\ref{5}(c) shows the relative
probability of that one particular set of data points to precede a high volatility day with probability proportional to circle radius.  Essentially, Fig.~\ref{5}(c) and (d) are heat maps with bubble size being used in lieu of color intensity.

Were there no next-day memory effect, all bubbles would be of equal size. However, in Fig.~\ref{5}(d) we find that the joint conditional probability for the top quintiles $P[g(t+1)=p|v_5(t),g_5(t)]$ is approximately three times the size of the unconditioned probability $P(g(t+1)=A)$, indicating that days with the top quintiles of both volatility and volume are overrepresented in the days preceding high volatility by a factor of three. In contrast, the probability for the bottom quintiles $P[g(t+1)=p|v_1(t),g_1(t)]$ is only half that of the unconditioned probability, meaning that days with the bottom quintiles in both volatility and volume are underrepresented in the days preceding high volatility by a factor of two. We compare this to the results yielded from the investigation in Fig.~\ref{5}(a) and (b), where the greatest overrepresentation by quintile is approximately only a factor of two.  This indicates that the volume and volatility combined are a more powerful predictor of upcoming high volatility than either volume or volatility alone. The variation of
results by both row and column also indicates that there is information potentially important for volatility prediction embedded into both
quantities.  We confirm this by applying a simple multiregression model predicting next-day volatility from either volatility alone or
volatility and volume together. We find an average 6\% increase in the $R^2$ value when volume is included.

By contrast, Fig.~\ref{5}(c) shows the distribution by quintile of volume and volatility for days preceding the very lowest volatility days.  The variation in bubble size is
considerably reduced compared to that of Fig.~\ref{5}(d), showing that days preceding low volatility are far more evenly distributed in volatility and volume.  Additionally, there are
no clear pronounced trends across row or column that would indicate a clear effect of either volume or volatility on the next day's volatility value.
\par
\section{Conclusion and further discussion}
We have examined the relationship between trading volume, volatility, and LMV using correlation and time-lagged correlation, conditional
probability distributions, as well as joint conditional probability analysis, and distribution fits we have proposed.
 We find that while the same-day correlation between the logarithmic volume and volatility is fairly week, the same day and time-lagged correlation between logarithmic volume and a quantity we introduce as ``local maximum volatility'' (LMV) are both very strong.
This finding may help explain the inconsistency between investors intuition about market stability during high volume days and the empirical fact that the relationship is not strong. Although it is essential that a trader understands the effects of trading volume \cite{KK,KKK}, the weak correlation coefficient ($\approx 0.2$) is unable to explain the importance of trading volume \cite{bize}. While humans often interpret correlations to be stronger than they are (i.e., illusory correlation \cite{Chapman}), in the case of volume-volatility correlations there are obvious mechanisms indicating their reality.  Thus, we further investigate and find out that through the strong correlation between volume and LMV,
a trader's interpretation may be justified. We believe LMV to be a more accurate representation of an investor's memory than the actual volatilities themselves. The cognitive bias in which humans disproportionately focus their attention on negative experiences and threats over positive experiences and aid is well-documented in cognitive psychology and termed the ``negativity bias'' \cite{Wason}, summarized by Baumeister et al.~\cite{Baumeister} as ``bad is stronger than good.''  The manifestation of negativity bias in trading in the form of volatility asymmetry---wherein negative price changes cause a market to become more volatile than positive price changes---has been observed in many different countries \cite{Engle2,Zakoian,Sentana,Jayasuriya,Tenenbaum}.  Thus our findings using LMV match the behavior of investors because LMV is a more important quantity when it comes to human perception.
An investor may thus be justified in having an negative attentional bias because (s)he does not know the next-day volatility level in
advance and must treat the ``risk of risk'' as the relevant quantity.

Our findings also indicate that high volatility tends to follow high trading volume, although low volatility is largely unaffected by volume.
Because we observe that high volatility strongly affects trading volume, we posit that volume can be used to predict future volatility, especially on days of highest volatility. Based on the new dynamics we provided and the empirical analysis, we determine the predictive ability of volume in estimating near-future high volatility. Our analysis shows
that volume is as useful in predicting future volatility as volatility itself and using both volume and volatility in the prediction is better than using
 either of them alone.  Further, we have introduced the functional form that gives the tail of the volume-conditional volatility distribution and shown that
 the distribution is unified across wide ranges of volumes when viewed in scaled units making the abscissa the volatility divided by the volume.  Thus,
 we are able to explain not only why high volatility tends to occur with large volume, but also to what extent the latter effects the former.

This work is supported by the grant ``Econophysics and Complex Networks'' (R-144-000-313-133) and we thank F. Ling and B. Podobnik for their constructive suggestions.


\begin{thebibliography}{99}
\bibitem{KK}
http://www.learn-stock-options-trading.com/stock-volume.html
\bibitem{KKK}
http://www.stocktradingtogo.com/2007/05/09/volume-interpretation-with-stock-charts/
\bibitem{Ane}
Ane T, Geman H (2000) Order flow, transaction clock, and normality of asset returns, Journal of Finance. 55: 2259-2284.
\bibitem{Karpoff}
Karpoff JM (1987) The relation between price changes and trading volume: A survey, J Financ Quant Anal, 22:109-126.
\bibitem{Crouch1}
Crouch RL (1970) A Nonlinear Test of the Random-Walk Hypothesis. American Economic Review 60(1):199-202.

\bibitem{Lamoureux}
Lamoureux CG,  Lastrapes W (1990) Heteroskedasticity in Stock Return Data: Volume versus GARCH Effects.
 The Journal of Finance 45:221-229.
\bibitem{Plerou2} 
 Plerou V, Gopikrishnan P, Gabaix X, Amaral L, Stanley HE (2001) Price fluctuations, market activity and trading volume.
Quantitative Finance 1: 262-269.
\bibitem{Ying} 
Ying C (1966) Stock Market Prices and Volumes of Sales.
Econometrica 34: 676-685.
\bibitem{Rogalski} 
 Rogalski RJ (1978) The Dependence of Prices and Volume.
Rev Econ Stat 60:268-274.
\bibitem{Tauchen1}
Tauchen G, Pitts M (1983 )The Price Variability-Volume Relationship on Speculative Markets.
Econometrica 51:485-505.
\bibitem{Whaley}
 Stephan JA, Whaley RE (1990) Intraday Price Change and Trading Volume Relations in the Stock and Stock Option Markets.
The Journal of Finance 45(1):191-220.
\bibitem{Gabaix1}
Gabaix X., Gopikrishnan P, Plerou V, Stanley HE (2003) A Theory of Power-Law Distributions in Financial Market Fluctuations. Nature 423:267-270.
\bibitem{Lobato}%
Lobato IN, Velasco C (2000) Long memory in stock-market trading volume.
J. Bus. Econom. Statist  18:410-427.
\bibitem{Brailsford}
Brailsford TJ (1996) The empirical relationship between trading volume, returns and volatility.
Accounting \& Finance 36:89–111.
\bibitem{Brooks}
 Brooks C (1998) Predicting Stock Index Volatility: Does Volume Help?
Journal of Forecasting 17:59–80 (1998).
\bibitem{Clark}
Clark PK (1973) A Subordinated Stochastic Process Model with Finite Variance for Speculative Prices.
Econometrica 41(1):135-155.
\bibitem{Koulakiotis}
Koulakiotis A, Dasilas A, Molyneux P  (2007) Does Trading Volume Influence GARCH Effects? – Some evidence from the Greek Market with Special Reference to Banking Sector.
Investment Management and Financial Innovations 4(3): 33-38.
\bibitem{Sharma}
Sharma JL, Mougoue M, Kamath R, (1996) Heteroscedasticity in stock market indicator return data: volume versus GARCH effects.
Applied Financial Economics 6(4): 337-342.
\bibitem{Gillemot}
Gillemot L, Farmer JD, Lillo F (2006) There's More to Volatility than Volume,
 Quantitative Finance, 6(5): 371-384.
 \bibitem{Chambers}
Chambers JM, Hastie TJ (1992)
Statistical Models in S (Wadsworth \& Brooks/Cole Computer Science)
Wadsworth \& Brooks/Cole. CRC Press.
\bibitem{Wang}
Wang F, Yamasaki Y, Havlin S, Stanley HE (2010) Statistical Regularities of Equity Market Activity
arXiv:0911.4258,
\bibitem{Wang2}
Wang F, Shieh SJ, Havlin S, Stanley HE (2009) Statistical Analysis of the Overnight and Daytime Return,
 Phys. Rev. E 79: 056109.
\bibitem{Podobnik2009}
Podobnik B, Horvatic D, Petersen AM, Stanley HE (2009) Quantitative Relations between Risk, Return and Firm Size,
EPL 85: 50003.
\bibitem{Podobnik}
Podobnik B, Horvatic D, Petersen AM, Stanley HE (2009)
Proc. Natl. Acad. Sci. USA 06: 22079 (2009)
\bibitem{Yamasaki}
K. Yamasaki, L. Muchnik, S. Havlin, A. Bunde, and H. E. Stanley, Proc. Natl. Acad. Sci. U.S.A. {\bf 102}, 9424 (2005).
\bibitem{Gallant}
A.R. Gallant, P. E. Rossi and G. Tauchen,
Rev Finance Stud {\bf 5}, 199 (1992).
 \bibitem{zheng}
Z. Zheng, B. Podobnik, L. Feng, and B. Li,
Nature Scientific Reports {\bf 2}, 888 (2012)
\bibitem{Stanley}
M. H. R. Stanley, L. A. N. Amaral, S. V. Buldyrev, S. Havlin, H. Leschhorn, P. Maass, M. A. Salinger, and H.~E. Stanley, Nature {\bf 379}, 804 (1996).
\bibitem{Chapman}
L. Chapman,
Journal of Verbal Learning and Verbal Behavior {\bf 6 (1)}, 151 (1967).
\bibitem{Wason}
P. C. Wason,
The Quarterly Journal of Experimental Psychology  {\bf 11} (2), 92 (1959)
\bibitem{Baumeister}
R. Baumeister, E. Bratslavsky, C. Finkenauer, and K. Vohs,
Review of General Psychology {\bf 5 (4)}, 323 (2001).
\bibitem{Engle2}R.~F. Engle and V. Ding, J. Financ. {\bf 48}, 1749 (1993).
\bibitem{Zakoian}J.~M. Zakoian, J. Econ. Dyn. Control {\bf 18}, 931 (1994).
\bibitem{Sentana}E. Sentana, Rev. Econ. Stud. {\bf 62}, 639 (1995).
\bibitem{Jayasuriya}S.~A. Jayasuriya and R. Rossiter, J. Int. Finance and Eco. {\bf 8}, 11 (2008).
\bibitem{Tenenbaum}J. Tenenbaum, D. Horvatic, S.~C. Bajic, B. Pehlivanovic, B. Podobnik, and H.~E. Stanley. Phys. Rev. E {\bf 82} (2010).

\bibitem{bize} 
http://www.bized.co.uk/timeweb/crunching/crunch\_relate\_expl.htm


\end{thebibliography}
\end{document}